\documentclass[aps,prb,twocolumn,superscriptaddress]{revtex4-1}

\usepackage{graphicx}
\usepackage{dcolumn}
\usepackage{bm}
\usepackage[normalem]{ulem}

\begin{document}

\title{Phase transitions of H$_{\bm 2}$ 
adsorbed on the surface of single carbon nanotubes
}  

\author{M.C. Gordillo}
\affiliation{Departamento de Sistemas F\'{\i}sicos, Qu\'{\i}micos 
y Naturales, Facultad de Ciencias Experimentales, Universidad Pablo de
Olavide, Carretera de Utrera, km 1. 41013 Sevilla, Spain}
\author{J. Boronat}
\affiliation{Departament de F\'{\i}sica i Enginyeria Nuclear, 
Universitat Polit\`ecnica de Catalunya, 
Campus Nord B4-B5, 08034 Barcelona, Spain}

\date{\today}

\begin{abstract}
By means of Diffusion Monte Carlo calculations, we obtained the complete
phase diagrams   of H$_2$ adsorbed on the outer surface of isolated
armchair carbon nanotubes of radii ranging from 3.42 to 10.85 \AA. We only
considered density ranges corresponding to the filling of the first
adsorption layer in these curved structures.  In all cases, the zero-temperature
ground state
was found to be an incommensurate solid, except in the widest tube, in
which the structure with lowest energy is an analogous of the $\sqrt{3}
\times \sqrt{3}$ phase found in planar substrates. Those incommensurate
solids result form the arrangement of the hydrogen  molecules in
circumferences whose plane is perpendicular to the main axis of the carbon
nanotube.  For each tube, there is only one of such phases stable in the 
density range considered, except in the case of the (5,5) and (6,6) tubes,
in which two of these incommensurate   solids are separated by novel first
order phase transitions. 
\end{abstract}
 
\pacs{67.25.dp,05.30.Jp,87.80.bd,68.90.+g}

\maketitle

\section{Introduction}
A carbon nanotube is a cylindrical structure~\cite{Iijima} that can be
thought as  the result of the wrapping up of a single graphene 
sheet~\cite{science2004,pnas2005} over itself. In the same way that
\textbf{,} in the past \textbf{,} 
graphene sheets were always found stacked to form graphite,  carbon
nanotubes were always found forming nanotube bundles, structures in
which the only surfaces available for adsorption were the exposed parts of
the tubes in the fringes of the bundle~\cite{talapatra,prl2006,prb2007}.
The adsorbed phases in a planar structure     such as graphite
are quite different from the ones observed on a highly patterned substrate
like the  external surface of a nanotube bundle. This can be seen by
comparing the  results on a bundle of carbon nanotubes and graphite for
$^4$He \cite{glyde,prl2008} (bundles) \cite{grey2,prl2009} (graphite),
H$_2$ \cite{vilchesjltp1,vilchesjltp2} (bundles) \cite{frei1,frei3,prb2010}
(graphite), and Ne \cite{talapatra2,vilchesne} (bundles) and
\cite{tiby,colebook} (graphite). Similar to the opportunities for
adsorption that the achievement of a single graphene
sheet offers,  a recent experimental work shows the proper technique to
suspend a single carbon
nanotube and study the phases of different gases (Ar,Kr) adsorbed
on its
external surface~\cite{vilchesscience}. In this latter work, the results 
indicated that  the
phase diagrams were similar to those of  the same gases on graphite, but
with phase transitions shifted to higher pressures. In the
present work, we have studied
the adsorption behavior of  H$_2$ on a series of armchair
((n,n)) carbon nanotubes with different diameters in order to see if a 
quantum gas would exhibit novel phases when adsorbed on curved surfaces.
The carbon nanotubes considered, together  with their indexes and radii,
are given in Table I. The most noticeable results driven from our
calculations are: \textit{i}), the existence of a solid-solid phase
transition for (5,5) and (6,6) nanotubes, and \textit{ii}), the stability
of the curved version of the $\sqrt{3} \times \sqrt{3}$ commensurate phase
only for the widest nanotube studied (16,16).   

\section{Method}
Our calculations were performed with the diffusion Monte Carlo (DMC)
method. This well-established technique allows us  to solve exactly the 
Schroedinger equation within some statistical errors to
obtain the ground state energy of a system of interacting
bosons~\cite{boro94}.  
This is exactly our case, since in its lowest energy state
($para$-H$_2$) the hydrogen molecule behaves as a boson. To apply
the DMC algorithm, we need the potentials  to describe  the H$_2$-H$_2$
interaction (we chose the Silvera and Goldman expression~\cite{silvera}, a
standard model for calculations involving $para$-H$_2$), and   the C-H$_2$ one.
For the latter, we used a Lennard-Jones potential with
parameters taken from  
Ref. \onlinecite{coleh2}, as used previously to describe the phase
diagram of H$_2$ adsorbed on graphene \cite{prb2010}.  All the individual C-H$_2$
pairs were considered, i.e., we took into consideration the corrugation effects
due to the real structure of the nanotube.  

The last ingredient needed to perform a calculation in a DMC scheme is
a trial wave function. It regulates the Monte Carlo importance sampling, and 
can be considered as a variational approximation to the exact description of 
the system.
In this work, we used a trial wave function formed by the product of two terms. 
The first one is 
\begin{eqnarray}
\lefteqn{ \Phi({\bf r}_1, {\bf r}_2, \ldots, {\bf r}_N)  =  \prod_{i<j} \exp \left[-\frac{1}{2}
\left( \frac{b_{{\text H}_2{\text -}{\text H}_2}}{r_{ij}} \right)^5
\right]}  \label{trial1} \\ 
& & \times \prod_i  \prod_J \exp \left[ -\frac{1}{2} \left( \frac{b_{{\text
C}{\text -}{\text 
H}_2}}{r_{iJ}} \right)^5 \right]  
\prod_i \exp (-a (r_i-r_0)^2) \nonumber \ , 
\end{eqnarray}  
where ${\bf r}_1$, ${\bf r}_2$, \ldots, ${\bf r}_N$ are the coordinates of
the H$_2$ molecules   and ${\bf r}_J$ the position of the carbon atoms in
the nanotube. The first term in Eq. (\ref{trial1})  is a two-body Jastrow
function depending on the H$_2$ intermolecular distances $r_{ij}$,  while
the second one has the same meaning but for each $r_{iJ}$
(C-H$_2$). 
Finally, the third term is a product of one-body Gaussians with
$r_i=\sqrt{x_i^2+y_i^2}$,  that depend on
the distance of each H$_2$ to the center of the tube ($r_0$).

\begin{table}
\caption{Armchair carbon nanotubes considered in this work,
together with  their tube radii ($r_t$) and the most probable distance for
an adsorbed hydrogen molecule to  the center of the cylinder ($r_0$).
In the last column, the
adsorption energy of a single H$_2$ molecule ($e_0$) is
reported.   
}
\begin{tabular}{cccc} \hline
Tube & $r_t$ (\AA) & $r_0$ (\AA) & e$_0$ (K) \\ \hline
(5,5) & 3.42 & 6.36 &  -320.6 $\pm$ 0.1  \\
(6,6) & 4.10 & 7.05 &  -329.0  $\pm$ 0.1  \\
(7,7) & 4.75 & 7.70 &  -335.5  $\pm$ 0.1  \\
(8,8) & 5.45 & 8.39 &  -341.0 $\pm$ 0.1 \\
(10,10) & 6.80 & 9.75 &  -349.2 $\pm$  0.2 \\
(12,12) & 8.14 & 11.10 & -353.7  $\pm$ 0.2 \\
(14,14) & 9.49 & 12.46 & -356.5 $\pm$  0.1  \\
(16,16) & 10.85 & 13.80 & -357.2 $\pm$  0.2  \\
 \hline
\end{tabular}
\end{table}

Equation \ref{trial1} is appropriate for describing liquid phases, but if the 
system is a solid, we need to multiply  the trial wave function
above (\ref{trial1}) by  
\begin{equation}
 \prod_i \exp \left\{ -c \left[ (x_i-x_{\text{site}})^2+ (y_i - y_{\text{site}})^2 + 
(z_i - z_{\text{site}})^2 \right] \right\}  \ ,
\label{trialsol}
\end{equation}
whose purpose is to limit the location of the hydrogen molecules to regions
close to the   $x_{\text{site}},y_{\text{site}},z_{\text{site}}$
regularly
distributed coordinates of a solid arrangement. The variational parameters
appearing in the whole trial wave function ($b_{{\text H}_2\text{-H}_2}$, 
$b_{\text{C-H}_2}$, $a$ and $c$) were fixed to be the same values than the
ones for H$_2$ adsorbed on graphene~\cite{prb2010},  after having checked that 
there were no appreciable
differences when variationally optimized for the (5,5) tube, i.e.   
$b_{{\text H}_2\text{-H}_2} = 3.195$ \AA, $b_{\text{C-H}_2}$ =  2.3 \AA and
$a$ = 3.06 \AA$^{-2}$.  Parameter $c$ was varied with density in the same way than
for graphene.  The only remaining parameter, $r_0$, was optimized
independently for each tube; the results are  listed in Table I.

\section{Results}

Table I contains also the binding energies ($e_0$) of a single H$_2$
molecule on top of the series of nanotubes considered in this work. We
observe, quite reasonably, that this energy increases with the radius of
the cylinder. However, even the value for the (16,16) tube is quite far
from the result on a flat graphene sheet (-431.79 $\pm$ 0.06 K~\cite{prb2010}), 
due to the surface curvature, that distorts the C-H$_2$ distances with respect to those
of a flat graphene sheet.  

Our aim in the present work is to describe the possible phases of H$_2$ 
adsorbed on the surface of
different (n,n) nanotubes. To do  so, we considered  the curved
counterparts of all the commensurate solid phases found experimentally for 
most of the quantum gases ($^4$He, H$_2$ and D$_2$) on graphite~\cite{grey2,frei1,frei3,w}, 
which included the curved version of a $\sqrt{3}
\times \sqrt{3}$ phase,    perfectly possible in these (n,n) tube
substrates~\cite{vilchesscience}.  For the incommensurate phases, we tried structures
similar to the triangular phases found in graphene, but wrapped up around
to form hydrogen cylinders of radius $r_0$ (see Table I). One of these
structures is shown in Fig. 1, which displays the projections on a plane of
the H$_2$ site locations  for an incommensurate solid  wrapped around a (5,5)
tube. Here, the $r$ coordinate represents the hydrogen positions on a
circumference of radius $r_0$ = 6.36 \AA, while the $z$ axis is chosen
parallel to the main axis of the tube.   One can see that the solid is
built by locating five H$_2$ molecules in circumferences on planes perpendicular to $z$, and
rotating the molecules in neighboring circumferences  half the distance between
H$_2$ molecules. We defined a phase by the number of hydrogen molecules in
one of such circumferences, and varied their density by changing the distance between them. 
Alternatively, this structure can be thought as the result of having
five helices of pitch 17 \AA
wrapped around the tube, each one of them including 10 molecules per turn of the helix. 

\begin{figure}
\begin{center}
\includegraphics[width=0.8\linewidth]{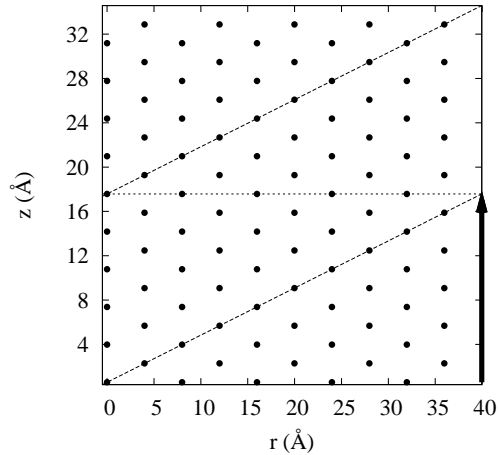}
\caption{Projection on a flat surface of an incommensurate structure corresponding to an areal density of
0.076 \AA$^{-2}$  to be wrapped around a (5,5) tube. It can be described
as an set of five intertwined helices whose pitch (indicated by an arrow) is 17   
\AA.  Every turn of the helix contains
10 H$_2$ molecules.   
}
\end{center}
\label{fig1}
\end{figure}

\begin{figure}
\begin{center}
\includegraphics[width=0.8\linewidth]{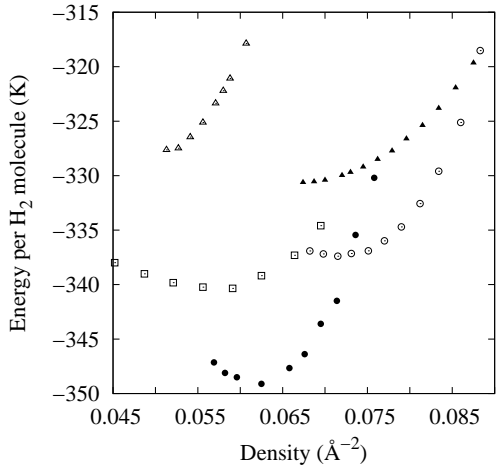}
\caption{Energies per H$_2$ molecule for different adsorbed phases on a
(5,5) tube. We display here the liquid (open squares), the different
incommensurate solids: five-in-a-row (solid circles), six-in-a row   (open
circles), four-in-a-row (open triangles) and seven-in-a-row (full
triangles). 
}
\end{center}
\label{fig2}
\end{figure}

DMC results for the equations of state corresponding to 
different phases  of H$_2$ adsorbed on a (5,5) tube are
shown in Fig. 2. Open squares indicate a liquid phase (obtained by
considering as a trial function only Eq. (\ref{trial1})), while the circles
represent solid incommensurate phases with five (full circles) and  six
(open circles) H$_2$ molecules per row. We display with triangles results for
a  four molecules per row (open), and seven molecules per row (full)
arrangements. For all the areal densities considered (up to 0.0937
\AA$^{-2}$, the experimental density for a second layer
promotion             in planar graphite \cite{colebook}), these last two
phases are unstable (of higher energy) with respect to the first two. All
the alternative commensurate structures are also unstable, as can be seen
in Table II and III. In that table, we also display their areal densities,
different for different tubes, even though the pattern of  the adsorbed
hydrogen molecules is the same. This is due to the fact that the areal
densities depend on $r_0$, while these phases are registered with respect
to structures whose dimensions depend on $r_t$.   Other solid
incommensurate structures, such as helices with different number of
molecules per turn but with the same pitch, i.e., more or less H$_2$ molecules on top of the dashed lines
in Fig. 1,   have energies greater than the phases represented by the
circles  in Fig. 2. For instance, a structure with eleven molecules per
turn, instead of the ten displayed in  Fig. 1, has a minimum energy of
-336.0 $\pm$ 0.1 K, for a density of 0.062 $\pm$ 0.003 \AA$^{-2}$. 
Therefore, the ground state of H$_2$ adsorbed in the outer surface of
a (5,5) carbon nanotube is an incommensurate solid with five H$_2$'s per
row.  Upon a density increase, there is a first-order solid-solid
phase transition to another
incommensurate solid, similar to the first one but with six atoms per row. 
The equilibrium densities for
both structures at the transition are obtained from a double
tangent Maxwell construction: 
0.0685  $\pm$ 0.0001 \AA$^{-2}$
(five molecules per row, energy per hydrogen molecule -345.3 $\pm$ 0.1 K) 
and 0.0795  $\pm$ 0.0001 \AA$^{-2}$ (six intertwined helices, with energy
per hydrogen molecule -334.7 $\pm$ 0.1 K).

\squeezetable
\begin{table*}
\caption{Energies per particle ($e_b$) and equilibrium densities
($\rho$) for different phases proposed  in the literature~\cite{grey2}
for quantum gases on graphite, when adsorbed in tubes of different radii.
The liquid and incommensurate helical densities are the values
corresponding to the minimum energies obtained by means of third-degree
polynomial fits to curves of the type displayed in Fig. 1. The rest
correspond to exact densities. Error bars within parenthesis represent
the uncertainty of the last figure shown.
}
\begin{tabular}{ccccccccccc} \hline
Phase   & & liquid &  & 2/5 & & 3/7 & & $\sqrt{3} \times \sqrt{3}$  & & helical incommensurate \\ \hline
Tube    & $\rho$ (\AA$^{-2}$)  & $e_b$ (K) & $\rho$  (\AA$^{-2}$)  & $e_b$ (K) & 
       $\rho$  (\AA$^{-2}$)  & $e_b$ (K) & $\rho$  (\AA$^{-2}$)  & $e_b$ (K) & 
       $\rho$  (\AA$^{-2}$)  & $e_b$ (K)  
      \\ \hline
(5,5) & 0.0574(4) & -340.3(1) & 0.0407 & -333.7(2) & 0.0436 & -325.4(2) &  0.0339 & -329.03(4) & 0.0621(3) & -349.0(2)  \\
(6,6) & 0.0586(3) & -349.70(9)& 0.0441 & -344.4(1) & 0.0472 & -334.3(2) &  0.0367 & -339.52(6) & 0.0673(1) & -360.34(4)   \\
(7,7) & 0.0615(6) & -356.2(2) & 0.0471 & -352.28(7)& 0.0504 & -339.7(2) &  0.0392 & -347.66(3) & 0.0716(1) & -367.89(5) \\
(8,8) & 0.0611(7) & -360.3(2) & 0.0494 & -358.9(1) & 0.0529 & -343.4(2) &  0.0411 & -354.44(3) & 0.0742(1) & -372.43(3) \\
(10,10)& 0.0578(3)& -365.95(6)& 0.0531 & -367.11(8)& 0.0569 & -347.5(2) &  0.0443 & -364.12(5) & 0.0758(1) &  -375.6(1)  \\
(12,12)& 0.0565(4)& -369.1(1) & 0.0559 & -371.8(1) & 0.0600 & -346.3(3) &  0.0466 & -369.93(5) & 0.0703(2) & -375.96(5) \\
(14,14)& 0.0530(4)& -370.4(1) & 0.0581 & -373.5(1) & 0.0623 & -342.5(3) &  0.0485 & -373.10(8) & 0.0681(2) & -375.3(1) \\
(16,16)&0.0511(8)&  -367.57(9)& 0.0600 & -371.17(5)& 0.0643 & -333.8(3) &  0.0500 & -371.7(1) & 0.0660(1) & -371.4(1)    \\
 \hline
\end{tabular}
\end{table*}

\begin{figure}
\begin{center}
\includegraphics[width=0.8\linewidth]{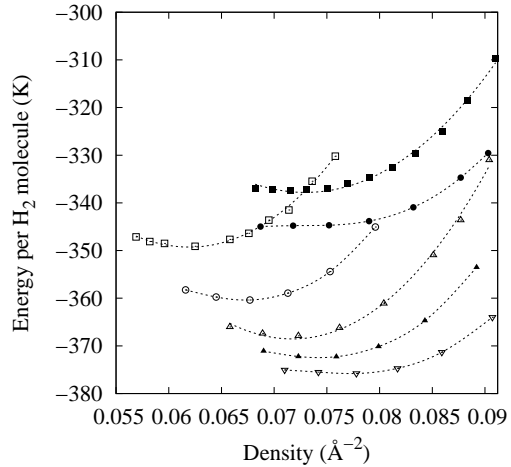}
\caption{Equation of state for the incommensurate solids adsorbed in carbon
nanotubes of increasing radii:  (5,5), (open squares, five-in-a-row solid;
full squares, six-in-a-row
arrangement); (6,6), (open circles, six-in-a-row structure; full circles, seven-in-a-row solid); (7,7),
(open triangles); (8,8), (full triangles); (10,10), (inverted triangles). 
See further explanation in the text. 
}
\end{center}
\label{fig3}
\end{figure}

The stable phases of H$_2$ on other (n,n) tubes, up to n = 14, are
similar to the (5,5) one. In all cases, the ground states are the same type
of incommensurate solids already described, the only difference being  the
number of intertwined helices (or molecules in the same circumference) forming the
structure. This number was found to be always equal to the index $n$ of the
nanotube. In Fig. 3, we show the equation of state for some of such solids,
from  the already displayed (5,5) case (open circles; five molecules in a
row), to the (10,10) one (inverted open triangles; ten molecules in a row),
going through the (6,6) (open circles), (7,7) (open triangles) and  (8,8)
(full triangles) tubes. In all cases the dashed lines are mere
guides-to-the-eye. Between the two narrowest cylinders and the
rest there is an important difference, though: the (6,6) tube exhibits a
first order phase transition between two incommensurate solids with six and
seven molecules per row while, while for the other tubes,  there is only
one 
stable incommensurate structure in the areal density range corresponding to
an adsorbed monolayer The zero-pressure densities and energies
corresponding to the ground state for all the tubes considered are given in
Table II. There, and in Table III, we can see that those incommensurate
solids are the truly ground states for the systems under consideration at
zero pressure, since their energies are lower than  the corresponding to
any of the commensurate structures on the same tubes. The only exception is
the (16,16) nanotube,  in which the ground state is the same
$\sqrt{3}
\times \sqrt{3}$ structure than in a flat surface. The stability limits 
for the two solids in the (6,6) tube are 0.0755 $\pm$ 0.0001 \AA$^{-2}$
($e_b$ = -338.2 $\pm$ 0.1 K) and 0.0855  $\pm$ 0.0001 \AA$^{-2}$ ($e_b$ =
-354.1 $\pm$ 0.1 K), also obtained by a Maxwell construction. The $1 \times
\sqrt{3}$ structure defined in Ref.~\onlinecite{colesolido} for (n,0)
nanotubes was found also to be unstable for all the tubes considered. For
instance, for the (16,16) tube, the density of this phase is 0.0750 
\AA$^{-2}$, with a binding energy of  -303.9 $\pm$ 0.6 K.

\begin{table}
\caption{Same than in Table II, but for two other commensurate structures
found experimentally for D$_2$ on graphite.  The dimensions of their unit
cells make them only possible for the tubes shown.   In all cases, they are
unstable with respect to the incommensurate structures (n$\leq$ 14) or to
the $\sqrt{3} \times \sqrt{3}$  one (n= 16). 
}
\begin{tabular}{ccc} \hline
 & 7/16\cite{corboz,w}  & \\ \hline
Tube  & $\rho$ (\AA$^{-2}$)  & $e_b$ (K) \\ \hline
(8,8)   & 0.0540  & -359.4 $\pm$  0.1  \\
(12,12) & 0.0612  & -371.6 $\pm$  0.1   \\
(16,16) & 0.0656  & -369.9 $\pm$ 0.1     \\ \hline
 & 31/75 ($\delta$ phase in H$_2$ on graphite \cite{w}) & \\ \hline
Tube  & $\rho$ & $e_b$ \\ \hline
(5,5)   & 0.0421  & -334.8  $\pm$  0.1 \\
(10,10) & 0.0549  & -367.0  $\pm$  0.1 \\
 \hline
\end{tabular}
\end{table}

\section{Concluding remarks}
Summarizing, we studied all the possible H$_2$ phases adsorbed on
the surface of armchair
carbon nanotubes ranging from (5,5) to (16,16). In all cases, but the last
one, we have found that the stable phases are incommensurate solids 
formed by 
molecules adsorbed on circumferences perpendicular to the main tube axis.
Where the curvature of the surface is more relevant, i.e., in the
narrowest tubes [(5,5),(6,6)], our results show the existence of a
solid-solid zero-temperature phase transition between two incommensurate structures.
The first commensurate solid phase, curved version of the well-known  
$\sqrt{3} \times \sqrt{3}$ phase, appears only when the radius of the
nanotube is big enough: a (16,16) tube for H$_2$.

\begin{acknowledgments}
We acknowledge partial financial support from the 
Junta de Andaluc\'{\i}a group PAI-205 and grant FQM-5985, DGI (Spain) Grants No.
 FIS2010-18356 and FIS2008-04403, and Generalitat de Catalunya 
 Grant No. 2009SGR-1003.
\end{acknowledgments}

\end{document}